\documentclass[noshowpacs,amsmath,
twocolumn,
superscriptaddress,
8pt,
aps,prb]{revtex4-1}  
\usepackage{setspace}
\usepackage{amsmath}
\usepackage{graphicx}
\usepackage{epsfig}
\usepackage{upgreek} 
\usepackage{verbatim}  
\usepackage{amsfonts}
\usepackage{amssymb}
\usepackage{textcomp}
\usepackage{epstopdf}
\usepackage{xcolor}
\usepackage{gensymb} 
\usepackage{upgreek} 
\DeclareGraphicsExtensions{.pdf,.eps,.png,.jpg,.mps}
\draft
\begin{document}
\title{Universal light-guiding geometry for high-nonlinear resonators\\having molecular-scale roughness}
\author{Dae-Gon Kim$^{1\dagger}$, Sangyoon Han$^{2\dagger}$, Joonhyuk Hwang$^{2}$, In Hwan Do$^{1}$, Dongin Jeong$^{1}$,\\Ji-Hun Lim$^{3}$, Yong-Hoon Lee$^{3}$, Muhan Choi$^{3}$, Yong-Hee Lee$^{2}$, Duk-Yong Choi$^{4*}$, Hansuek Lee$^{1,2*}$\\$^1$Graduate School of Nanoscience and Technology, Korea Advanced Institute of Science and Technology (KAIST), Daejeon 34141, Republic of Korea.\\$^2$Department of Physics, Korea Advanced Institute of Science and Technology (KAIST), Daejeon 34141, Republic of Korea.\\$^3$School of Electronics Engineering, Kyungpook National University, Daegu 41566, Republic of Korea.\\$^4$Laser Physics Centre, Research School of Physics,\\Australian National University, Canberra, ACT 2601, Australia.\\$^{\dagger}$These authors contributed equally to this work.\\$^{*}$duk.choi@anu.edu.au; hansuek@kaist.ac.kr}
\maketitle
\begin{figure*}[t!]
  \centering
  \includegraphics[trim = 0mm 0mm 0mm 0mm, width=15cm]{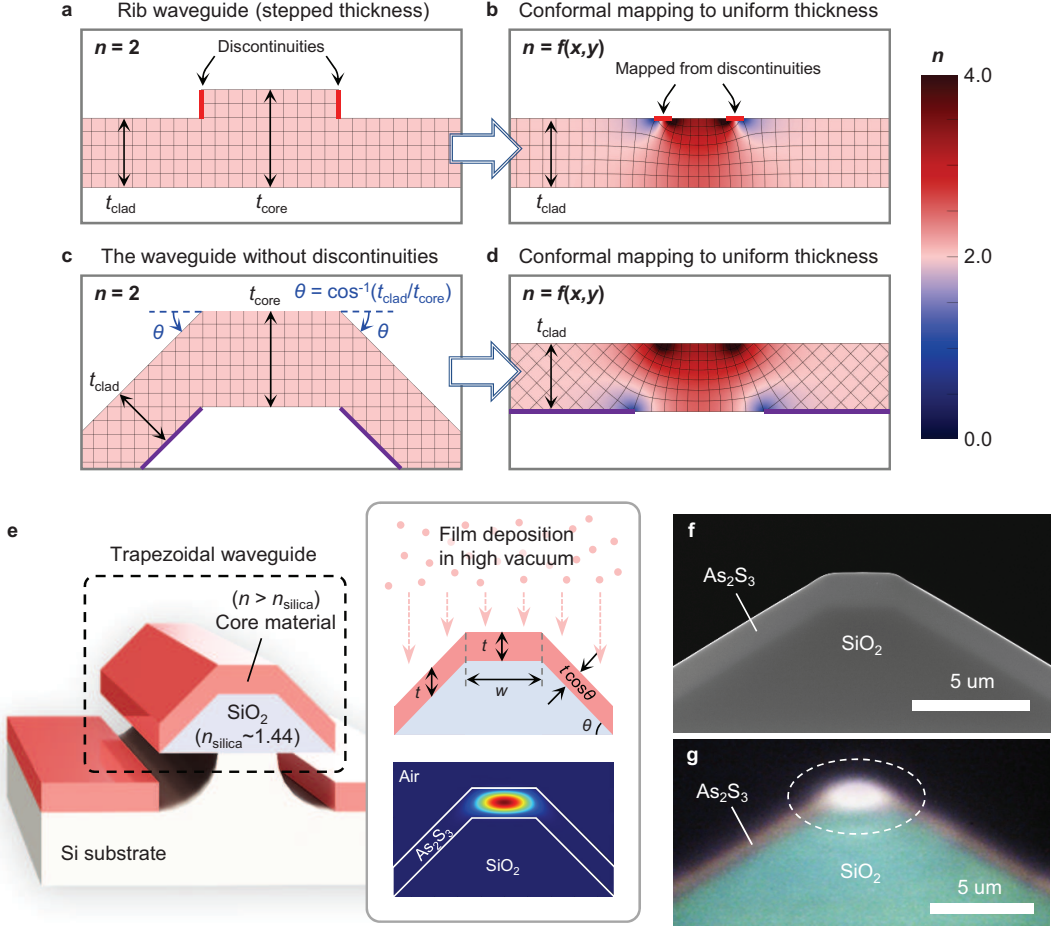}
  \caption{\textbf{Design concept of a trapezoidal waveguide and its implementation.} \textbf{a,c.} A cross-section of a typical rib waveguide and the proposed waveguide without the discontinuities, respectively. The refractive index of the material is set to 2 ($n=2$) without loss of generality. \textbf{b,d.} A refractive index profile of the flat waveguides conformally mapped from the rib and the proposed waveguide, respectively. \textbf{e.} Implementation of the proposed waveguide.  By depositing the core material on the SiO$_{2}$ platform structure, the waveguide having continuous junctions can be formed (inset, top). Simulated optical mode profile of the waveguide made of As$_{2}$S$_{3}$ ($n=2.43$) (inset, bottom). \textbf{f.} The cleaved cross-section of the fabricated waveguide observed by scanning electron microscope. \textbf{g.} An optical microscope image of the transmitted mode at the output facet. An IR camera image is superimposed with a visible camera image to clearly display the optical mode and waveguide structure.}
  \label{fig1}
\end{figure*}
\begin{figure*}[t!]
\centering
  \includegraphics[trim = 0mm 0mm 0mm 0mm, width=15cm]{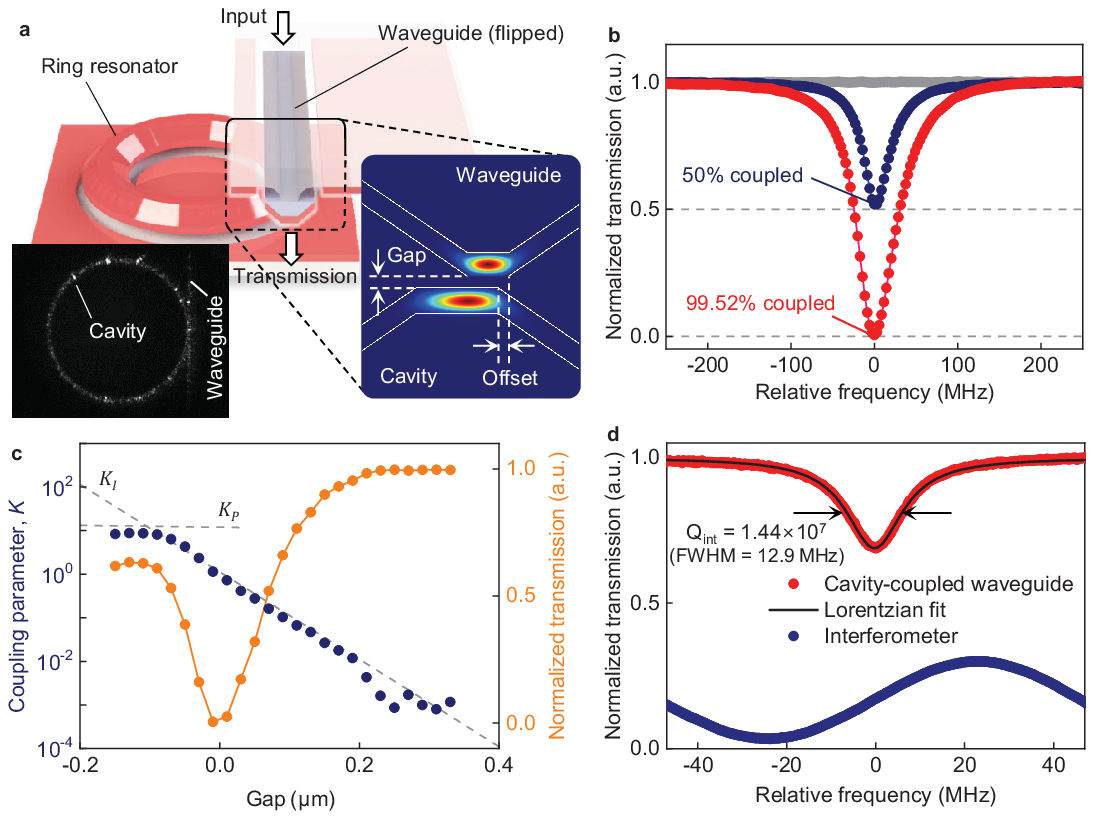} 
  \caption{\textbf{Optical coupling between the trapezoidal waveguide and the ring resonator.} \textbf{a.} Schematic image explaining the flip-chip coupling scheme. Optical modes in the waveguide and the resonator are evanescently coupled through their top surfaces. Coupling strength is controlled by adjusting the gap and the offset (inset, right). The left inset taken by IR camera shows the scattered light trace from the coupled chips. \textbf{b.} Transmission spectra with different coupling strengths. \textbf{c.} The gap versus coupling parameter $K$ (navy curve) converted from the normalized transmission $T$ (orange curve). The gap in the graph represents the relative value to the gap for critical coupling. \textbf{d.} Transmission spectrum for measuring Q-factor of the trapezoidal resonator (red curve) and its Lorentzian fitting (black line). Intrinsic Q factor of 14.4 million is obtained from the spectrum. The navy curve is the calibration spectrum from a fiber Mach-Zehnder interferometer.}
  \label{fig2}
\end{figure*}
\hspace{-10pt}\textbf{By providing an effective way to leverage nonlinear phenomena in chip-scale, high-Q optical resonators have induced the recent advances of on-chip photonics represented by micro-combs\cite{DODOS_Nat} and ultra-narrow linewidth lasers\cite{JBluementhal_Nat_Si3N4SBS}. These achievements mainly relying on Si, SiO$_{2}$, and Si$_{3}$N$_{4}$ are expected to be further improved by introducing new materials having higher nonlinearity. However, establishing fabrication processes to shape a new material into the resonator geometries having extremely smooth surfaces on a chip has been a challenging task. Here we describe a universal method to implement high-Q resonators with any materials which can be deposited in high vacuum. This approach, by which light-guiding cores having surface roughness in molecular-scale is automatically defined along the prepatterned platform structures during the deposition, is verified with As$_{2}$S$_{3}$, a typical chalcogenide glass of high-nonlinearity. The Q-factor of the developed resonator is 14.4 million approaching the loss of chalcogenide fibers\cite{FiberLoss_NRL}, which is measured in newly proposed tunable waveguide-to-resonator coupling scheme with high ideality. Lasing by stimulated Brillouin process is demonstrated with threshold power of 0.53 mW which is 100 times lower than the previous record based on chalcogenide glasses\cite{BEggleton_SBSdevice}. This approach paves the way for bringing various materials of distinguished virtues to the on-chip domain while keeping the loss performance comparable to that of bulk form.}\\
\indent
High-Q optical resonators, which had long been used for scientific researches such as cavity quantum electrodynamics\cite{aoki2006observation} and optomechanics\cite{kippenberg2008cavity}, have become an irreplaceable core of on-chip photonics with huge practical impacts. With dedicated efforts over a past decade, now, resonators can be fabricated on a chip with high Q-factor\cite{HLee_Nat_Photonics,DODOS_Nat,Lipson_Nat_battery}. Their geometry can be precisely designed and shaped to control the major properties such as free spectral range (FSR) and dispersion to satisfy the requirement for the efficient nonlinear process\cite{HLee_Nat_Photonics,KVahala_NatPhoton_DoubleWedge}. This advance has implanted versatile functions established in nonlinear optics to on-chip photonics while keeping its own virtues such as miniaturization, mass-productivity, and integration with other active and passive components. Optical frequency comb which revolutionized ultrafast science and metrology is being implemented on a chip based on Kerr nonlinearity\cite{DODOS_Nat,Lipson_Nat_battery}. A core part of ultra-narrow linewidth lasers\cite{JBluementhal_Nat_Si3N4SBS} and highly sensitive optical gyroscopes\cite{KVahala_arXiv_gyro} have been demonstrated in chip-scale by means of stimulated Brillouin scattering (SBS) process.\\
\indent
There are great motivations to explore various materials although the current achievements are entirely relying on Si, SiO$_{2}$, Si$_{3}$N$_{4}$, the primary materials for integrated photonics. Materials having higher nonlinearity give an opportunity to decrease pump power to the level required for portable devices\cite{BEggleton_Nat_Photonics,gQ_MLoncar_NatComm_LBO,gQ_KYvind_Optica_AlGaAS,JHu_NP_integrated}. Materials having extended transmission windows can spread this advance in the near-infrared to neighboring wavelength ranges\cite{BEggleton_Nat_Photonics}. Accordingly, a lot of attention is paid to develop on-chip high-Q resonators with other promising materials. The most critical part of these attempts is to develop the etch process customized for the particular nature of each material to attain extremely smooth surface since the Q-factor is mainly limited by surface-scattering loss. This challenge which recurs whenever introducing new materials impedes further development of the technology beyond the current state which is limited by the property of the primary materials. To completely address this issue, we propose a universal approach by which high-Q resonators, or more generally light-guiding structures with extremely low loss, can be defined on a chip without the following etch process.\\
\indent
The indispensability of etching or subtractive processes for waveguide fabrications is related to the discontinuity of the waveguide geometry. Transformation optics \cite{xu2015conformal} provide insight into the origin of the discontinuity. Forming a light-guiding structure can be generally understood as the practical implementation of a core, the domain having a relatively higher refractive index than a surrounding medium. For the most common case of using a single material, a core needs to be defined by controlling the thickness distribution as a rib waveguide shown in Fig. 1a. The refractive index profile of the equivalent flat waveguide in Fig. 1b, which is quasi-conformally transformed from the rib\cite{chang2010design}, clearly shows higher index around the region corresponding to the thickest part of the original structure. The mapping is performed to result in the equivalent structure having a uniform thickness while keeping the geometry of outer thinner parts, namely a cladding, unchanged. It converts the thickness variation to the index distribution. It is noteworthy that, in conventional approaches, discontinuity or sudden change of the film geometry necessarily occurs around the junction between the thick core and thin cladding as marked by the red lines in Fig. 1a. Therefore, to practically fabricate such discontinuities, it is unavoidable to perform direct etch processes or any indirect processes such as lift-off which cause rough surfaces. Moreover, these rough surfaces strongly interact with the optical modes since they are mapped near the highest index region where most of fields are confined.\\
\indent
To resolve this problem, we introduce a new light-guiding structure without the discontinuities as shown in Fig. 1c. To continuously connect the claddings having a thickness of $t_{\textrm{clad}}$ to the core having a thickness of $t_{\textrm{core}}$, the claddings are rotated by $\theta$ which satisfies $\cos\theta=t_{\textrm{clad}}/t_{\textrm{core}} (-90^{\circ}<\theta<90^{\circ})$). The transformed geometry in Fig. 1d obviously shows the formation of a high index region around the top flat section. This waveguide having continuous junctions can be realized simply by depositing the core material on the bottom platform structure made of SiO$_{2}$ without subsequent etch processes. As shown in Fig. 1e, the cross-section of the platform structure is a trapezoid having both side slopes corresponding to the characteristic profile formed by wet etching\cite{HLee_Nat_Photonics}. The deposition under high vacuum results in the uniform film thickness of $t_{\textrm{core}}$ along the deposition direction over the whole platform. Therefore, the film thickness on the side slope along the normal direction from the material interface decreases to $t_{\textrm{core}}\cos\theta$ which satisfies the continuous junction condition for arbitrary values of $t_{\textrm{core}}$ and $\theta$.\\
\indent
The design of trapezoidal waveguides involves intrinsic immunity for surface roughness of platform structures. In contrast to the rib waveguide, the etched surfaces of the silica platform marked by the purple lines in Fig. 1c are well separated from the high index region in the transformed geometry. Therefore, optical modes are mainly guided by the surfaces having molecular roughness defined by film deposition and the unetched oxide interface. Moreover, the roughness of the wet-etched silica surface can be suppressed below 1 nm as demonstrated for ultra-high-Q wedge resonators\cite{HLee_Nat_Photonics}.\\
\indent
This universal approach applicable to any materials with a higher refractive index than SiO$_{2}$ which can be directionally deposited on a wafer is experimentally verified with chalcogenide glass. Chalcogenide glass has attracted great attention due to its distinguished high nonlinearity as well as high transparency in the mid-infrared wavelength range\cite{BEggleton_Nat_Photonics,OBang_Nat_Photonics,FiberLoss_NRL}. However, its inherent irregular nature related to the variation of stoichiometry and bond structures tends to cause the characteristic grainy rough surface by nonuniform etch rate in nanoscale\cite{DYChoi_JAP}. As a result, the loss performance of on-chip chalcogenide devices\cite{BEggleton_LowlossWG,JHu_OL_highQ} has lagged far behind that of optical fiber form\cite{FiberLoss_NRL} or theoretical estimation\cite{TheoLoss_referedbyNRL} despite continuous and dedicated efforts over a decade to tame this particular nature.\\
\indent
The cross-sectional image of the waveguide fabricated with As$_{2}$S$_{3}$, a typical chalcogenide glass, is shown in Fig. 1f. It is obvious that the film is uniformly deposited on the SiO$_{2}$ platform by means of thermal evaporation with a large mean free path (around 500 m at 10$^{-7}$ torr). The surface roughness of the SiO$_{2}$ platform and As$_{2}$S$_{3}$ film is suppressed below 1 nm, confirmed by atomic force microscopy. The fabrication procedure and details are provided in the method section.\\
\indent
The confinement of the optical mode in the top-flat region is experimentally verified by observing the transmitted mode profile from the cleaved output facet of a straight waveguide with a near-infrared camera as shown in Fig. 1g. This mode profile of 1550 nm wavelength corresponds well to the intensity distribution of the fundamental TE mode simulated by Lumerical Mode Solutions. The coupling loss for end-fire coupling with a lensed fiber is 3.5 dB, attained without serious engineering to match the mode profiles for higher coupling efficiency.\\
\indent
Although on-chip ring resonators can be easily implemented with trapezoidal waveguides, proper coupling scheme is required to efficiently access the modes in the resonators. The efficient coupling for chalcogenide resonators has been a critical issue because of its relatively higher refractive index which cannot be matched by a SiO$_{2}$ tapered fiber, a well-developed coupling tool in wide use\cite{Tbirks_OL_taperfiber,KVahala_PRL_taperfiber}. Although there have been considerable efforts to fabricate tapered fibers with chalcogenide glass\cite{lecaplain2016mid}, its mechanical instability and difficulty in tapering process hinder wide-spread of this technique.\\
\begin{figure*}[t!]
  \centering
  \includegraphics[trim = 0mm 0mm 0mm 0mm, width=15cm]{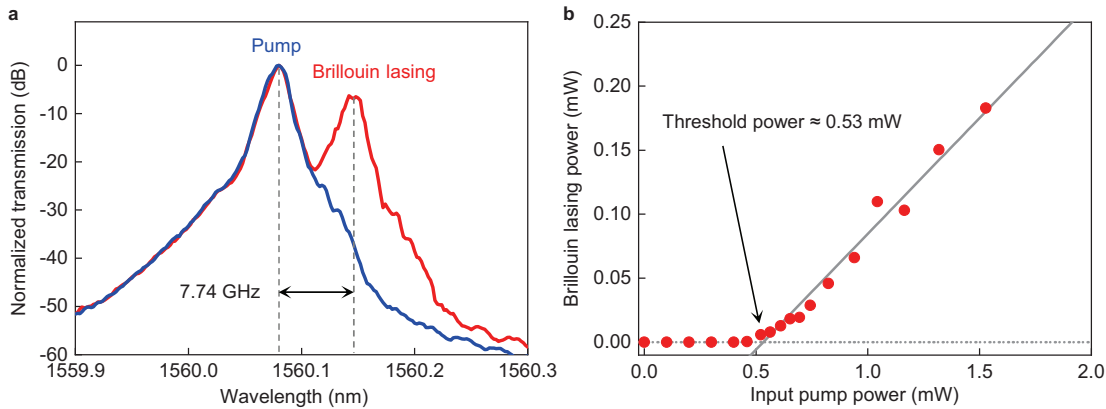}
  \caption{\textbf{Brillouin laser with a sub-mW threshold power using a high-Q resonator.} \textbf{a.} Optical spectra obtained by an optical spectrum analyzer with pump powers below threshold (blue line) and above threshold (red line). The Brillouin shift of 7.74 GHz is measured by an electrical spectrum analyzer. \textbf{b.} Input pump power versus Brillouin lasing power. A lasing threshold power of 0.53 mW is obtained by a linear fitting of the data (gray line).}
  \label{fig3}
\end{figure*}
\indent
As a general solution to couple with resonators made of any materials, a new flip-chip coupling scheme is developed as shown in Fig. 2a, where an on-chip bus waveguide is flipped over and coupled to a resonator on the other chip. Since both guiding structures have a similar trapezoidal cross-section, the effective refractive indices of their modes match well. The scattered trace of the mode in the coupled chips observed by IR camera which can see structures through the flipped-over chip is shown in the left inset of Fig. 2a. The coupling strength can be adjusted as needed by controlling the gap and offset (Fig. 2a right inset) with a piezo stage. As confirmed in Fig. 2b, the resonance mode in under-coupled and critical-coupled condition clearly fits with Lorentzian function without a vestige of Fano resonance which is typically observed for a multimodal bus waveguide.\\
\indent
For further investigation of the coupling junction performance, the coupling parameter $K$ and the ideality $I$ have been calculated from the transmission on resonance measured with varying the gap distance between the waveguide and resonator\cite{KVahala_PRL_ideality,TKippenberg_PRA_ideality}. Since the coupling parameter can be adjusted equal to or larger than 1 as shown in Fig. 2c, it is confirmed that over-coupling condition beyond the critical coupling is achievable. The coupling performance is quantified by the ideality of the junction that is the degree to which the junction between a bus waveguide and resonator behave as a true single-mode coupler. As described in detail in the method section, it is confirmed that the ideality value is 0.923 in the critical coupling condition. Here, zero of the ideality means no coupling into the desired mode and unity stands for the perfect single-mode coupler. Therefore, the measured ideality shows that the proposed coupling scheme allows selective and efficient access to the targeted mode without significant power loss or addition of redundant complexity by multimode interaction.\\
\indent
Q-factor, the most important feature, is measured to evaluate the performance of the developed resonators. The dimensions of the resonator are 1.3 \(\upmu\)m of As$_{2}$S$_{3}$ film thickness, 10 \(\upmu\)m of top width, 30$^{\circ}$ of slope angle, and 4.94 mm of diameter. By scanning the laser frequency, the full width half maximum (FWHM) of the resonance is measured, which is calibrated by the sinusoidal reference signal of 100.63 MHz from a fiber Mach-Zehnder interferometer as shown in Fig. 2d. The maximum Q-factor attained from FWHM of 12.9 MHz is $1.44\cdot10^{7}$, which is more than 10 times larger than the previous record of the on-chip resonators made of chalcogenide glass\cite{JHu_OL_highQ}. It should be emphasized that the waveguide loss of 2.77 dB/m converted from the measured Q-factor approaches 0.9 dB/m, the best record of the loss of chalcogenide glass fiber at conventional telecom wavelength\cite{FiberLoss_NRL}. This implies that the scattering loss induced by the trapezoidal waveguide is well suppressed to the level of the material loss of chalcogenide glass which is dominated by material absorption and Rayleigh scattering by nanoscale refractive index fluctuation.\\
\indent
Among the possible applications exploiting the large optical nonlinearity of chalcogenide glass, we demonstrate on-chip lasing based on the SBS process. The dimensions of the resonator are the same as the previous one for Q-measurement, which is designed to place the SBS process in the cavity. FSR of its TE fundamental mode is 7.74 GHz corresponding to the Brillouin phonon frequency in As$_{2}$S$_{3}$ at 1560 nm wavelength. Characterization of the Brillouin lasing signal is performed in the measurement setup similar to the previous studies\cite{BEggleton_SBSreview,KVahala_arXiv_gyro}. Figure 3a shows the measured optical spectrum of the back-reflected pump and the Brillouin lasing signal (1$^{st}$ Stokes wave). The beating frequency between them is 7.74 GHz which corresponds to a Brillouin frequency shift of As$_{2}$S$_{3}$ fibers\cite{sanghera2010nonlinear}. In this figure, pump intensity is higher than the Brillouin lasing signal because around 17.5\% of the pump is back-reflected at the input facet of the coupled waveguide due to the large difference of refractive index between the air and chalcogenide glass.\\
\indent
Significant reduction of a lasing threshold is achieved by the enhanced Q-factor and Brillouin gain coefficient of As$_{2}$S$_{3}$ that is tens of times larger than those of SiO$_{2}$ and Si$_{3}$N$_{4}$. The lasing signal power depending on input pump power is measured while the pump laser is locked to the resonant mode as plotted in Fig. 3b. It shows a typical lasing curve with a slope efficiency of 18\%. The lasing threshold is 530 \(\upmu\)W which is around 100 and 25 times lower than the previous record of the on-chip Brillouin lasers made of chalcogenide and nitride, respectively. It is remarkable that Q-factor is inverse-quadratically related to the threshold power of not only the Brillouin lasing but also the other nonlinear processes such as Raman lasing and four wave mixing.\\
\indent
Additional features proving the potential of this approach are briefly described as follows. It is numerically demonstrated that dispersion can be geometrically controlled by adjusting the structure of the trapezoidal waveguides. Considering the ease of stacking multilayer cores in thickness resolution of nanometer-scale, precise dispersion control through material combination is also achievable. For mid-IR application, excess material absorption caused by SiO$_{2}$ platform can be completely prevented by depositing a bottom cladding layer before the core deposition.\\
\indent
In summary, we have presented the universal approach to define the light-guiding structure of extremely smooth surfaces on a chip without etching the core materials as well as the flip-chip coupling scheme to evanescently couple the developed resonators with the high coupling ideality. By applying this scheme to chalcogenide glass, the threshold of the on-chip Brillouin laser is reduced by about 100 times than that of the previous chalcogenide devices, by means of the enhanced Q-factor which is close to the loss of state-of-the-art optical fibers made of the same material. This universal method can be applied to bring the other materials having distinguished merits to the on-chip domain while keeping the loss performance comparable to that of bulk form.\\

\hspace{-10pt}\textbf{Acknowledgments} This work was supported by Samsung Research Funding \& Incubation Center of Samsung Electronics under Project Number SRFC-IT1801-03.\\

\hspace{-10pt}\textbf{Author Contributions} All authors made important contributions to the work.\\

\hspace{-10pt}\textbf{Competing interests} The authors declare no competing financial interests.\\

\hspace{-10pt}\textbf{Correspondence and requests for materials} should be addressed to D.-Y.C. and H.L.\\
\clearpage
\section*{METHODS}
\hspace{-10pt}\textbf{Formation of SiO$_{2}$ trapezoidal platform.} Extended Data Fig. 1a shows details for forming SiO$_{2}$ trapezoidal platform. First, a layer of SiO$_{2}$ is thermally grown with a thickness of 8 \(\upmu\)m on a prime-grade (100) Si wafer in a furnace system (Tytan Mini 1800, TYSTAR). The SiO$_{2}$ layer is treated with hexamethyldisilazane (HMDS) vapor prime process to modify surface adhesion property. To define the SiO$_{2}$ trapezoidal platform, we then pattern photoresists on the SiO$_{2}$ layer with a mask aligner lithography system (MA6 Mask Aligner, Karl Suss). The patterned photoresist is used as an etch mask for the following wet-etching process. The wet etching of SiO$_{2}$ layer is done with buffered hydrofluoric acid (BHF) solution. The slope angle of the platform can be controlled in the range from 5$^{\circ}$ to 60$^{\circ}$ according to the adhesion property between the photoresist and the SiO$_{2}$ layer, which is mainly controlled with the HMDS treatment step. In addition, by repeating the steps through the photolithography and wet-etching, the slope angle can reach larger values than the normal range attained by a single process\cite{KVahala_NatPhoton_DoubleWedge}. The details for realizing the smooth etched surface with desired slope angle is described in the previous work\cite{HLee_Nat_Photonics}. After the wet-etching, a cleaning process with organic solvents is performed to remove the photoresist.\\

\hspace{-10pt}\textbf{Isolation from Si substrate.} Once the SiO$_{2}$ trapezoidal platform is formed, we isotropically etch the Si substrate under the platform with XeF$_{2}$ gas (XeF$_{2}$ etching system, Teraleader Co. Ltd.) as shown in Extended Data Fig. 1b. By having this step, the optical field at the trapezoidal waveguide is perfectly isolated from the Si substrate to prevent leakage to the substrate. Alternatively, the optical field can be isolated from the substrate by growing a thermal oxide film on the Si substrate exposed by the wet-etching (Extended Data Fig. 1c).\\

\hspace{-10pt}\textbf{Film deposition.} Once the isolation step is done, a material deposition step follows (Extended Data Fig. 1 f,g). The material we use for the light-guiding core in this paper is As$_{2}$S$_{3}$ (AMTIR-6, Amorphous Materials Inc.), and the thickness is 1.3 \(\upmu\)m. The material is deposited with thermal evaporation process, and the deposition direction is normal to the wafer substrate. After the deposition, 2 nm-thick Al$_{2}$O$_{3}$ layer is formed on the surface by atomic layer deposition for surface passivation. A two-step annealing process (thermal and light annealing) reported in the previous work is performed to bring the As$_{2}$S$_{3}$ film close to the bulk state\cite{DChoi_PP_ChGannealing}. To estimate surface roughness of the waveguide, we analyzed a topography of the waveguide surfaces using atomic force microscopy (AFM). On the bare SiO$_{2}$ structure, the $rms$ roughness on the top and the wedge was 0.3 and 0.8 nm, respectively. Whereas on the as-deposited As$_{2}$S$_{3}$ surface, the roughness on the top and the wedge was 0.3 and 0.6 nm, respectively.\\

\hspace{-10pt}\textbf{Calculating coupling ideality, $I$.} Details for calculating coupling parameters and ideality can be found in the previous studies\cite{KVahala_PRL_ideality,TKippenberg_PRA_ideality}, but are summarized here for the convenience of the reader. Coupling parameter $K$ can be calculated from transmission of the waveguide $T$ (coupled to the resonator at resonant wavelength) using the following equation $K=(1\pm\sqrt{T})/(1\mp\sqrt{T})$ (upper signs for the over-coupled regime and lower signs for the under-coupled regime). $K$ can be decomposed into an intrinsic contribution $K_{\textrm{I}}$ and a parasitic contribution $K_{\textrm{P}}$, which stand for the coupling of a resonator mode to a fundamental mode and the other various parasitic components including higher order modes, respectively, so that 1/$K$  =  1/$K_{\textrm{I}}$+1/$K_{\textrm{P}}$. $K_{\textrm{I}}$ and $K_{\textrm{P}}$ can be found by asymptotically fitting gap versus $K$ graph (in logarithmic scale) with two lines (Fig. 2c). The line with smaller slope indicates $K_{\textrm{I}}$ and the line with larger slope indicates $K_{\textrm{P}}$. Coupling ideality, $I$, can be calculated from $K_{\textrm{P}}$ by using the following equation I =  1/(1+$K_{\textrm{P}}^{-1}$). The parasitic coupling parameter which is clearly revealed in the gap less than -0.1 \(\upmu\)m in the figure gives the ideality value of 0.923 in the critical coupling condition.\\

\hspace{-10pt}\textbf{Data availability.} The data sets generated and/or analysed during the current study are available from the corresponding authors on reasonable request.\\

\bibliographystyle{naturemag}
\bibliography{ms}
\renewcommand\thefigure{\arabic{figure}}
\setcounter{figure}{0}
\begin{figure*}[t!]
  \begin{center}
  \centerline{\includegraphics[trim = 0mm 0mm 0mm 0mm, width=14cm]{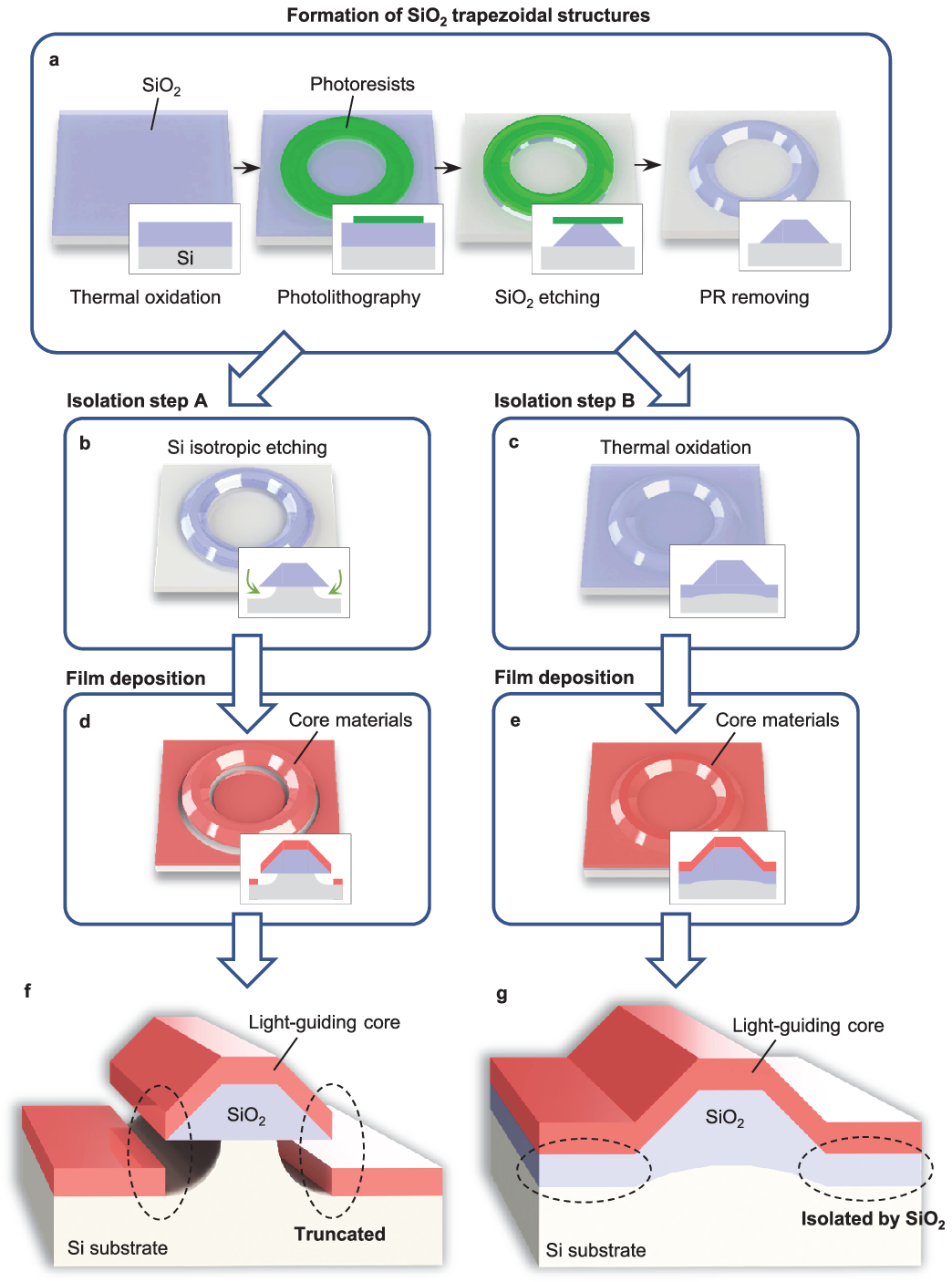}}
  \caption{\textbf{Fabrication procedure to realize the high-Q resonator (or the trapezoidal waveguide) on a chip.} \textbf{a.} Sequence to form SiO$_{2}$ trapezoidal structures. \textbf{b.} Isotropic XeF$_{2}$ etching of a Si substrate to isolate an optical field from the substrate. \textbf{c.} Additional thermal oxidation as an alternative to the isolation step. \textbf{d.} Film deposition of core materials on the XeF$_{2}$ etched structure. \textbf{e.} Film deposition of core materials on the additionally oxidized structure. \textbf{f.} The XeF$_{2}$-etched high-Q resonator in which the film of the core material is truncated at the end of the wedge structure. \textbf{g.} The additionally oxidized high-Q resonator in which the film of the core material is isolated from Si substrate by the SiO$_{2}$ sub-layer.}
  \label{EDfig1}
  \end{center}
\end{figure*}
\indent
\end{document}